\documentstyle[12pt,epsf]{article}
\begin{document}
\baselineskip 16pt plus 2pt minus 2pt

\newcommand{\beq}{\begin{equation}}
\newcommand{\eeq}{\end{equation}}
\newcommand{\beqa}{\begin{eqnarray}}
\newcommand{\eeqa}{\end{eqnarray}}
\newcommand{\dida}[1]{/ \!\!\! #1}
\renewcommand{\Im}{\mbox{\sl{Im}}}
\renewcommand{\Re}{\mbox{\sl{Re}}}
\def\simge{\hspace*{0.2em}\raisebox{0.5ex}{$>$}
     \hspace{-0.8em}\raisebox{-0.3em}{$\sim$}\hspace*{0.2em}}
\def\simle{\hspace*{0.2em}\raisebox{0.5ex}{$<$}
     \hspace{-0.8em}\raisebox{-0.3em}{$\sim$}\hspace*{0.2em}}

\begin{titlepage}

%\hfill{Version 07/19/99 pm}

\hfill{TRI-PP-99-027}

\vspace{1.0cm}

\begin{center}
%\vspace{.5cm}
{\large 
{\bf Singlet Interacting Neutrinos in the Extended Zee Model and 
Solar Neutrino Transformation}}\\

\vspace{1.2cm}
G. C. McLaughlin\footnote{email gail@lin04.triumf.ca} and
J. N. Ng\footnote{email: misery@triumf.ca}

\vspace{0.8cm}

TRIUMF, 4004 Wesbrook Mall, Vancouver, BC, Canada V6T 2A3\\[0.4cm]
\end{center}

\vspace{1cm}

\begin{abstract}
We study the impact of Standard Model singlet neutrinos on
 neutrino flavor transformation.  We focus
on an extension of the Zee model which includes singlet
neutrinos, and find that the best limits on the interactions of the
singlet neutrinos come from astrophysical phenomena.  Singlet
neutrino-electron scattering will impact 
both the mattter enhanced flavor transformation
potential as well as detector cross sections.
If electron neutrino - singlet neutrino oscillations are
responsible for the
solar neutrino anomaly, then the limit on the singlet neutrino
interaction strength is of order of the weak interaction scale.
Zee model modification of $\nu_\tau - e$ scattering also 
impacts solar neutrino transformation, although this 
interaction is more tightly constrained. \\[0.3cm]
{\em PACS}:  13.15+g, 12.60-i, 26.5+t\\ 
\end{abstract}
        
\vspace{2cm}
\vfill
\end{titlepage}

%\section{Introduction}

Recent data on atmospheric, solar and accelerator neutrino experiments 
\cite{lep} have indicated
that neutrinos have a finite mass and they exhibit flavor oscillation phenomena. 
Although the data has yet to be confirmed they point to the possibility that the 
three known active neutrinos are insufficient to account for all observations. 
The mass squared differences required to fit the data are: 
$10^{-11} \, {\rm eV}^2\leq \delta m^2_s \leq 10^{-5} \, {\rm eV}^2$
 , $\delta m^2_a \simeq 10^{-3} \, {\rm eV}^2$ and 
$0.2 \, {\rm eV}^2 \leq \delta m^2_{\mathrm LSND}\leq 2 \, {\rm eV}^2$ where the
subscripts $s$, and $a$ refer to solar and atmospheric oscillations respectively. 
At least one light standard model singlet fermion is required to reconcile
the data with the standard two by two neutrino mixing explanations. 

In most studies of a SM singlet neutrino one assumes it has no interactions 
other than gravitation at low energies and its role is to supply a mass for the active neutrinos.
An example of a singlet neutrino is a right-handed neutrino of the Dirac type. 
Another example comes from SO(10) GUT model where 
the  heavy SM singlet is a Majorana neutrino. In this model, the heavy neutrino
has a mass of $10^{12-14}$ GeV and
gives the active neutrinos a small mass via the seesaw mechanism. Its large mass 
prevents it from having direct experimental consequences at current available accelerator 
energies. 

Recently there has been renewed interest in models of radiatively generated 
neutrino masses with new physics occurring at or not much above the weak scale. 
An example is the Zee model which generates neutrino mass with the
same mechanism that produces lepton number violation \cite{zee}.  
The crucial ingredient is a lepton flavor changing SU(2) singlet scalar 
with U(1) hypercharge Y=2.     It is
straight forward to include a light SM neutrino in the model
\cite{fy,mn}, so as to accommodate 
all of the anomalous neutrino data.  We observe that now this light singlet
neutrino now has an interaction with a strength determined by the mass of 
the Zee scalar and its Yukawa couplings to the SM leptons. A detailed
phenomenological study of the current limits 
on these parameters are given in Ref \cite{mn}.

This paper is concerned with the effects of SM singlet neutrinos that have 
interactions as weak as or weaker than the normal weak interactions. 
These interactions can arise from a spin 1 particle exchange as in extended gauge 
models or a spin 0 boson exchange  such as the charged scalar in the Zee model.
The interactions may be weak due to small couplings and/ or a large mass of the
mediating particle.  One advantage of the Zee model is that it has relatively few parameters. Most
are constrained by terrestrial experiments. This makes it a economical model
for the study of singlet interacting neutrinos (SINs) on astrophysical phenomenon. Since there is
now a large body of solar neutrino data with more to come, we look
into the effects of SINs on some of the proposed solutions to the solar neutrino problem. 
For the purpose of this paper we can ignore the mixing of the Zee scalar with the
requisite two Higgs doublets without loss of generality.

SINs can affect the study of the neutrino fluxes from the sun in two ways.
They can impact on matter enhanced flavor transformation and secondly they alter the 
neutrino detector cross sections since they now interact with electrons albeit very weakly.
The interactions of the singlet will come into
play in any scenario which involves the Mikheyev-Smirnov-Wolfenstein (MSW) mechanism
\cite{msw} of neutrino transformation between an active neutrino and a singlet.  We 
illustrate both effects with the small angle solution to the solar neutrino problem.
If electron neutrinos transform to 
singlet neutrinos in the sun, the singlet neutrino coupling in the Zee model is 
most tightly constrained by the neutrino-electron scattering data at 
SuperKamiokande \cite{superk}.   We also show that a model which produces neutrino mass, 
such as the Zee model,
may provide additional interactions for the 
active neutrinos which will in principle occur in active-active 
transformation scenarios and detector cross sections.  
As we will demonstrate however, these parameters are 
constrained by terrestrial experiments to have a much smaller effect on 
neutrino flavor transformation. 

We begin by looking at the general equation governing the 
transformation of electron neutrinos
into another type of neutrino, where x = $\mu$,$\tau$, or s, and  s is the singlet 
neutrino, in a matter environment. This is given by
\begin{equation}
i\hbar \frac{\partial}{\partial r} \left[\begin{array}{cc} \Psi_e(r)
\\ \\ \Psi_x(r) \end{array}\right] = \left[\begin{array}{cc}
\varphi(r)    
& \sqrt{\Lambda} + V_{ex}(r)\\ \\ \sqrt{\Lambda}  + V_{xe}(r)
&  - \varphi(r)  
\end{array}\right]
\left[\begin{array}{cc} \Psi_e(r) \\ \\ \Psi_x(r)
  \end{array}\right]\,,
\label{eq:msw}
\end{equation}
where
\begin{equation}
  \label{eq:phie} \varphi(r) = \frac{1}{4 E} \left(
 (V_e(r) -V_x(r)) E - \delta m^2
  \cos{2\theta_v} \right)
\end{equation}
\begin{equation}
  \label{eq:ve} V_e(r) \equiv  \pm
 2 \sqrt{2}\ G_F \left[
  N_e^-(r) - N_e^+(r) - \frac{N_n(r)}{2} \right] 
\end{equation}

In these equations
\begin{equation}
  \label{eq:lambda}
  \sqrt{\Lambda} = \frac{\delta m^2}{4 E}\sin{2\theta_v},
\end{equation}
$\delta m^2 \equiv m_2^2 - m_1^2$ is the vacuum mass-squared
splitting, $\theta_v$ is the vacuum mixing angle, $G_F$ is the Fermi
constant, and $N_e^-(r)$, $N_e^+(r)$, and $N_n(r)$ are the number
density of electrons, positrons, and neutrons respectively in the
medium.  In the formulas given here, upper signs (in this case plus)
 correspond to the mixing of
neutrinos while the lower signs (in this case minus), 
correspond to the mixing of antineutrinos.  
The potential 
$\varphi_x(r)$ will have a standard model value and an additional term 
due to the extra interactions introduced by the Zee model. Other models 
which produce neutrino masses such as R-parity violating supersymmetric
models may have similar interactions.  Focusing on the Zee model, the charged scalar, $h^-$, 
is constrained by 
experiments at LEP to have a mass  $(M_h > 100 \, {\rm GeV})$. This in turn induces the 
following four fermion effective Lagrangian:
\begin{equation}
{\cal L} =  {|f_{12}|^2 \over  2 M_h^2} \bar{\nu}_\mu 
\gamma_\mu \nu_{\mu L} \bar{e} \gamma^\mu e_L.
\end{equation}
which governs low energy $\nu_\mu e$ scattering.
Note that this term has the opposite sign as the SM charged current 
interaction and is a prediction of the model. 
This term has to be added to the standard model muon neutrino MSW potential
as follows:
\begin{equation}
\label{eq:phimu} V_\mu(r) \equiv   \mp
 2 \sqrt{2}\ G_F \left[
  \delta (N_{e^-}(r) - N_{e^+}(r)) + { N_n(r) \over 2} \right] 
\end{equation}
where $\delta = \sqrt{2} |f_{12}|^2  / (8 M_h^2 G_F)$.
The  coupling constant $|f_{12}|^2$ is constrained by the
measurements of the lifetime of the muon to be 
$|f_{12}|^2 / ({M_h \over 100 {\rm GeV}})^2 < .0015$ (see \cite{mn}).  
Therefore, $ \delta  < .002$.  Similarly, 
$\nu_e \leftrightarrow \nu_\tau$ oscillations
will be influenced by the Zee model via additional $\nu_\tau - e$ scattering contribution to
the matter potential. The
potential takes the form of Eq. \ref{eq:phimu} with  
the replacement of $V_\mu$ by  $V_\tau$ and 
$\delta = \sqrt{2} |f_{13}|^2  / (8 M_h^2 G_F)$.  
The limit on  $|f_{13}|^2$ is derived from LEP and SLC
\cite{slc}
measurements of the leptonic vector and 
axial vector couplings in Z decay.  For a scale mass of 
$M_h = 800 \, {\rm GeV}$; the limit is $\delta < 0.1$. Clearly the effect
can be much larger for $\nu_e - \nu_\tau$ oscillation. This upper limit
gives rise to a maximum change in the MSW potential of about 10\%.
  
The extension of the Zee model which includes a singlet neutrino produces 
neutrino-electron scattering terms which, after Fierz transformation,
 have the form 
\begin{equation}
{\cal L} =  {|g_1|^2 \over 2 M_h^2} \bar{\nu}_R \gamma^\mu \nu_R \bar{e}_R 
\gamma_\mu e_R.
\end{equation}
Here, $g_1$ is the Zee model coupling between the singlet neutrino, the 
right handed electron and the scalar.  This produces a singlet 
neutrino MSW potential of
\begin{equation}
\label{eq:phis} V_s(r) =  \mp
   2 \sqrt{2}\ G_F 
  \beta  (N_{e^-}(r) - N_{e^+}(r))  
\end{equation}
where 
\begin{equation}
\label{eq:beta}
 \beta = \left( { \sqrt{2} |g_1|^2 \over 8 M_h^2 G_F} \right).  
\end{equation}
The best terrestrial
limit on $\beta$ comes from the leptonic right handed coupling to the Z \cite{mn}.
 The singlet neutrino does not couple directly to the Z boson, it only makes
a contribution to the decay through a correction at one loop order.  
The limit on ${|g_1|^2 /  2 M_h^2}$
is about $ 2 \times 10^{-4} \, {\rm GeV}^{-2}$, 
therefore $\beta < 6$. Other than this one loop effect we found no direct experimental bound
since this singlet neutrino can only enter in weak interaction processes via leptonic
mixing and no usable constraint is available. This in
principle allows the singlet-electron interaction to be larger
than the weak interaction. However, as we shall see below astrophysical considerations 
can provide tighter constraints. 

It is a characteristic of the Zee model that there are no interactions which convert electron
neutrinos directly to muon, tau or singlet neutrinos by way
of electron scattering mediated by the charged scalar.  Therefore,
in this model $V_{ex} = V_{xe} \approx 0$.

However, there are neutrino electron scattering terms in the
Zee model which convert muon (tau)  neutrinos to singlet neutrinos and
vice versa.  These are governed by the Lagrangian,
\begin{equation}
{\cal L} = - 
{ f_{12 (13)} g_1^* \over 2 M_h^2} \left( \bar{\nu}_R \nu_{\mu (\tau)} \bar{e}_R e_L -
{1 \over 4} \bar{\nu}_R \sigma^{\mu \nu} \nu_{\mu (\tau)} \bar{e}_R \sigma_{\mu \nu} e_L \right)
+ h.c. 
\end{equation}
In the forward scattering direction, 
the scalar term is proportional to neutrino mass and can therefore be
neglected.  The tensor term is proportional to electron spin and 
integrates to zero for unpolarized electrons.  Therefore, for most situations, 
$V_{\mu s (\tau s)}= V_{s \mu (s \tau)} =0$.  

Returning to the case of $\nu_e - \nu_s$ transition, 
the combined active-singlet neutrino transformation 
potentials can be cast in the form:
\begin{equation}
V_e - V_s = \pm { 3 G_F N_N(r) \over \sqrt{2}} 
\left[ \left( 1 + { 2 \over 3} \beta \right) Y_e - {1 \over 3} \right]
\end{equation} 
where $N_N$ is the total number density of nucleons. 
The electron fraction is defined as 
\begin{equation}
Y_e \equiv {N_{e^-}(r) - N_{e^+}(r) \over N_N }
\end{equation}
A non-zero $ \beta$ will have the effect of increasing the potential if the electron
fraction is greater than 1/3. In the sun for example, 
the electron fraction ranges from a value of about two thirds at the 
center to more than 0.85 in the outer layers.  The main effect of the 
new potential is to change the position at which a given neutrino undergoes the MSW 
resonance.  The position of the resonance is determined by the condition:
\begin{equation}
\label{eq:res}
 V_e - V_s =  \delta m^2  \cos{2\theta_v}. 
\end{equation}
Therefore, for given mixing parameters and a constant electron fraction, 
increasing $\beta$ causes the 
resonance position for a neutrino of energy E to shift to lower density. 

This scenario may be applied to several phenomenon, 
such as solar neutrinos and 
supernova neutrinos for the $\nu_e \leftrightarrow \nu_s$ or $\nu_e
\leftrightarrow \nu_{\mu,\tau}$ situation.  For the up-down asymmetry in the
atmospheric neutrino problem, the oscillations between muon neutrinos and either tau 
neutrinos or singlet may be analyzed in a similar manner, although we note
that in general there will be off-diagonal terms for 
$\nu_\mu \leftrightarrow \nu_\tau$ mixing.

Taking again the example of the sun, we see from Eq. \ref{eq:res} that
all neutrinos will pass through 
the resonance condition at a 
position that is further from the center of the sun as $\beta$ increases. 
For fixed $\delta m^2$ and $\sin^2 2  \theta_v$, low 
energy neutrinos that in the case of $\beta = 0$ did not encounter resonances
in the sun, will do so now if $\beta > 0$.  We illustrate this point
in Fig. \ref{fig:survival} where the survival probability for solar neutrinos is plotted for
two values of the parameter $\beta$.  This figure was produced by numerically 
integrating Eq (\ref{eq:msw}) for the singlet neutrino.  
For $\beta = 0$ the solution reduces to  
the small angle sterile neutrino oscillation 
solution to the solar neutrino problem, as in given in, for example
\cite{hata}.  It can be seen that 
increasing $\beta$ will cause a decrease in the number of low energy 
electron neutrinos coming from the sun.  

In contrast, the nonzero $ \beta$ has little effect on the high energy neutrinos.
The effective weak potential scale height 
\begin{equation}
L_V = \left| {d \ln (V_e - V_s) \over dr} \right|^{-1}
\end{equation}
which determines the survival probability at the resonance position 
remains fairly constant with small changes in  
$\beta$.  In fact for a fixed $Y_e$ and an exponential density profile
and a given neutrino energy, it can
be shown that the weak potential scale height remains constant
regardless of the
value of $\beta$. Plots similar to Figure \ref{fig:survival}
 can be drawn for $\nu_e \leftrightarrow 
\nu_{\mu,\tau}$ oscillations,
although nonzero $\delta$ will have a smaller impact on the survival probability 
as the constraints on $\delta$ are tighter.
    
It is also important to take into account the effect of a non-zero $\beta$ in
neutrino detectors.  In the Zee model, the neutrinos have only 
additional interactions with other 
leptons and not with the quarks due to the weak charge of the scalar. Hence,
 the radiochemical solar neutrino experiments such as SAGE and GALLAX
\cite{sage, gallax} {\em will not} be 
impacted by the additional interactions the singlet neutrino has. However, they will
register the low energy neutrino flux which depends on  $\beta$   
as explained before.  On the other hand, an experiment which 
detects
 neutrino-electron scattering, such as 
SuperKamiokande however, will have some 
portion of its signal coming from 
singlet neutrino-electron scattering if singlet neutrinos are present.  
The cross section for singlet neutrino electron scattering at first order 
is given by:
\begin{equation}
\label{eq:scross}
{d \sigma \over dT} =  {G_F^2 m_e \over 2 \pi} 4 \beta^2
\end{equation}
where T is the electron recoil energy.  In comparison,
the largest contribution to the standard model
$\nu_e -e$ scattering cross section approximately  given by:
\begin{equation}
\label{eq:ecross}
{d \sigma \over dT} =  {G_F^2 m_e \over 2 \pi} \left[ 
\left( 1+ 2 \sin^2 \theta_W \right)^2
+ (2 \sin^2 \theta_W)^2 \left( 1- {T \over E} \right)^2 + {\cal O} (m_e/E) \right].
\end{equation}
The theoretical
rate per electron recoil energy can be calculated by folding the survival 
probability for the electron neutrinos and the oscillation probability for 
sterile neutrinos with the cross sections and the fluxes of neutrinos.  
The detector rate may be estimated 
by folding the theoretical rate with an energy resolution function 
 as in Eq 4 of Ref \cite{plamen}.  

If $\nu_e \leftrightarrow \nu_s$
transformation is the solution to the solar neutrino problem, then it can be readily
seen that a constraint on $\beta$ comes from the overall number of events in 
Superkamiokande.  Various combinations
of mixing parameters and values of $\beta$ will produce different total count rate.  
If any significant mixing of electron neutrinos
and singlet neutrinos takes place in sun, and 
the flux predicted by the standard solar model 
\cite{ssm} is correct, 
then $\beta$ must be smaller than $\sim 1/2$.  This limit is much stronger than
the best limit from accelerator experiments of $\beta < 6$ and is
derived  from the extreme 
situation where all electron
neutrinos above 5 MeV are converted to singlet neutrinos.  

Figure \ref{fig:small} plots the ratio of rates with to without matter enhanced flavor 
transformation, using the shape  of the ${{^8\mathrm B}}$  neutrino
spectrum from  \cite{bahcall}.  Several
curves are plotted representing several mixing parameters and values of $\beta$.
It is seen that increasing the value of $\beta$ causes a flattening of the recoil
spectrum curve. It does not account for the upturn in event rate at high energy
observed in the SuperKamiokande data \cite{superk}.

Correctly reproducing the overall rates in all of the solar neutrino
experiments with a singlet neutrino - electron scattering interaction,
requires an adjustment to the usual sterile neutrino MSW mixing parameters.
For example if $\beta = 0.3$, then $\delta m^2$ must be increased by 
$\sim 30 \%$ (see Eq. \ref{eq:res}) in order to avoid reducing the fluxes
of pp neutrinos.  The mixing angle must be adjusted to take into account 
both the effect of the change in $\delta m^2$ on the survival probabilities 
of the ${{^8\mathrm B}}$ neutrinos and the effect of the nonzero singlet neutrino- electron 
scattering cross section in Kamiokande and Superkamiokande.  For $\beta = 0.3$,
$ \sin^2 2 \theta_v$ must be increased by $\sim 7 \%$.  The survival probability
and expected electron recoil spectrum for these parameters is shown by the dot-dashed
lines in Figures \ref{fig:survival} and \ref{fig:small}. 

We turn to the case of $\nu_\tau - \nu_e$ oscillations.  The $\nu_\tau - e$ 
scattering cross section can increase by a maximum of a factor of 2, depending
on the neutrino and electron energy,
since the Zee model amplitudes 
and the standard model amplitudes add coherently.  The 
scattering cross section in this case looks like
\begin{equation}
\label{eq:ecrossneut}
{d \sigma \over dT} =  {G_F^2 m_e \over 2 \pi} \left[ 
\left( 1 - 2 \sin^2 \theta_W  + 2 \delta\right )^2 
+ (2 \sin^2 \theta_W)^2 \left( 1- {T \over E} \right)^2 + 
{\cal O} (m_e/E) \right].
\end{equation}
For $\delta = 0$, this reduces to the standard model neutral current cross
section.  We illustrate the effect for matter enhanced flavor transformation
with parameters $\delta m^2 = 5.4 \times 10^{-6} \, {\rm eV}^2$ and 
$\sin^2 2 \theta_v = 6.3 \times 10^{-2}$.  In order to reproduce the 
observations in solar neutrino experiments
with nonzero $\delta$, the change in the
MSW potential will force a maximum increase in $\delta m^2$ of 10\%  above
the $\delta = 0$ solution.  For this change in $\delta m^2$ the   
 mixing parameter, $\sin^2 2 \theta_v$ retains approximately its original value
in order to take into account the both change in $\delta m^2$ and the
increase in scattering at the detector.
 
For comparison we consider the case of vacuum oscillations.  There
is no change in the survival probability of electron neutrinos due to
the singlet interactions.  However, the electron recoil spectrum will be effected.
The singlet neutrino - electron scattering  gives a signal similar to the 
neutral current scattering, although the size of the effect depends on the unknown magnitude 
of $\beta$.  Figure \ref{fig:vacuum}
 shows the electron recoil spectrum for a vacuum solution
of $\delta m^2 = 6.6 \times 10^{-11}$ and $\sin^2 2 \theta_v = 0.9$.  The active
neutrino solution is very similar to the singlet neutrino solution with $\beta = 0.3$.
Therefore in the vacuum case, as in the small angle case, 
it may be  difficult to differentiate
between singlet neutrino oscillations and active neutrino oscillations just by detecting
neutrino-electron scattering.

On the other hand, SNO \cite{SNO} may be able distinguish the active-singlet solution
from both the active-active oscillation and the active-sterile oscillation
through the combination of the three reactions:  (1) $\nu_e + d \rightarrow p + p + e^{-}$,
(2) $\nu_x + d \rightarrow p + n + \nu_x$ and neutrino-electron scattering
(3) $\nu_x + e^{-} \rightarrow \nu_x + e^{-}$.  In reaction 2, $\nu_x$ can be 
 $\nu_e$,
$\nu_\mu$, and $\nu_\tau$.  In reaction 3, $\nu_x$ includes the three active 
neutrinos and also the singlet neutrino.  The expected event rate  
per kilotonne year 
from the three reactions may be found in \cite{SNO}.  The standard solar model predicts
6500 events from reaction 1 above a 5 MeV threshold, and about 2200 will be seen for
vacuum $\nu_e \leftrightarrow \nu_\tau$ oscillations.  
These numbers remain roughly unchanged in the presence of 
a singlet or sterile neutrino, although there will be some variation depending
on the choice of mixing parameters.  The number of events for reaction two is 710 from the
standard solar model and remains
unchanged in the case of active-active oscillations, although it should be
reduced by about a factor of two in the case of active-sterile or active-singlet
oscillations, depending on the mixing parameters.  
From reaction 3, about 320 events are expected for
 active-active oscillations with about 78
coming from $\nu_{\mu(\tau)}$.  The number of neutrino-singlet interaction events
depends on the strength of the interaction $\beta$.  Therefore, if reaction
 2 indicates sterile or singlet neutrinos, for large $\beta$
reaction 3 in combination with 1 must be used to distinguish the two cases.       

In conclusion, we find that the extended Zee model with its new lepton number
violating interactions gives rise to new neutrino-electron scattering mechanisms.
The additional scattering occurs for muon, tau and singlet type neutrinos. 
It alters  the  small angle MSW $\nu_e - \nu_s$ and $\nu_e - \nu_\tau$
solar solution by shifting 
the resonance position for neutrinos of a given energy.  
Furthermore for it
increases the number of counts in water detectors 
and modifies the shape of the electron recoil 
spectrum produced from neutrino electron scattering.  Taking these into account
we have obtained a limit on singlet neutrino - electron interaction $\beta$  
which is an order of magnitude better that derived from  
current accelerator experiments.  With non-zero Zee model interactions
for the singlet and tau neutrinos, the apparent $\delta m^2$ as measured
by solar neutrino experiments can differ by as much as $\sim 50\%$ and 
$\sim 10\%$ respectively,
from that which would be measured by reactor experiments.         

This work is partially supported by a grant from the Natural Science and Engineering 
Council of Canada.

\newpage

\begin{figure}
%\epsfxsize=14cm
%\epsfxsize=14cm
%\centerline{\epsffile{fig1.ps}}
\caption{\label{fig:survival}
Survival probability for electron neutrinos produced near
the center of the sun is plotted against neutrino energy.  
The solid line is for electron neutrino-singlet neutrino
mixing parameters of 
$\delta m^2 = 4 \, \times \, 10^{-6} {\rm eV}^2$, $\sin^2 2 \theta_V = 0.01$ 
and a singlet neutrino interaction parameter, $\beta = 0$.  This
corresponds to the case of sterile neutrinos.  
The dashed line is for  $\delta m^2 = 4 \, \times \, 10^{-6} {\rm eV}^2$,
$\sin^2 2 \theta_v = 0.01$, $\beta = 0.3$ and shows that increasing the value
of $\beta$ causes less electron neutrinos to survive. 
The dot-dashed lines represents
$\delta m^2 = 5.2 \, \times \, 10^{-6} {\rm eV}^2$,
$\sin^2 2 \theta_v = 0.011$, $\beta = 0.3$.  The solid and dot-dashed lines
will produce very similar signals in current solar neutrino detectors.}
\end{figure}

\begin{figure}
%\epsfxsize=14cm
%\epsfxsize=14cm
%\centerline{\epsffile{fig2.ps}}
\caption{\label{fig:small}
The ratio of the number of electrons produced by neutrino
electron scattering
to the predicted number of
electrons using an undistorted $^{8}{\rm B}$ spectrum 
in a detector such as SuperKamiokande is shown.    
The solid, dashed and 
dot-dashed lines correspond to the same parameters as in Figure \ref{fig:survival}.  
Increasing $\beta$
causes a flattening of the curve and overall increase in the number of counts. 
}
\end{figure}

\begin{figure}
%\epsfxsize=14cm
%\epsfxsize=14cm
%\centerline{\epsffile{fig2.ps}}
\caption{\label{fig:vacuum}
Shows the same ratio is as in Figure \ref{fig:small} for vacuum oscillations.
The solid line corresponds to vacuum oscillations into active neutrinos,
where the dashed line corresponds vacuum oscillations into
singlet neutrinos with interaction parameter $\beta = 0.3$.}
\end{figure}

\newpage

\setcounter{figure}{0}
\begin{figure}
\epsfxsize=14cm
\centerline{\epsffile{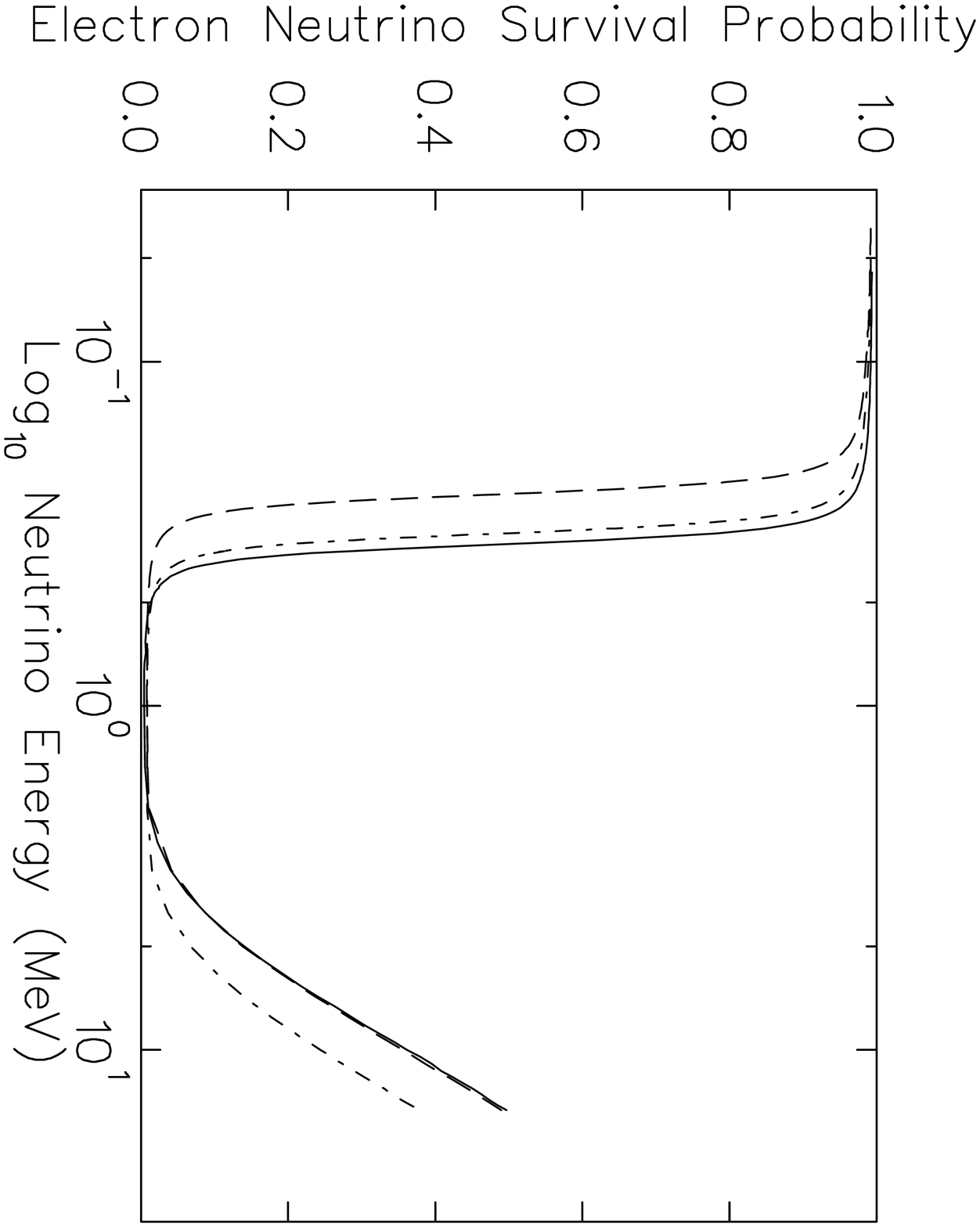}}
%\caption{}
\end{figure}

\begin{figure}
\epsfxsize=14cm
\centerline{\epsffile{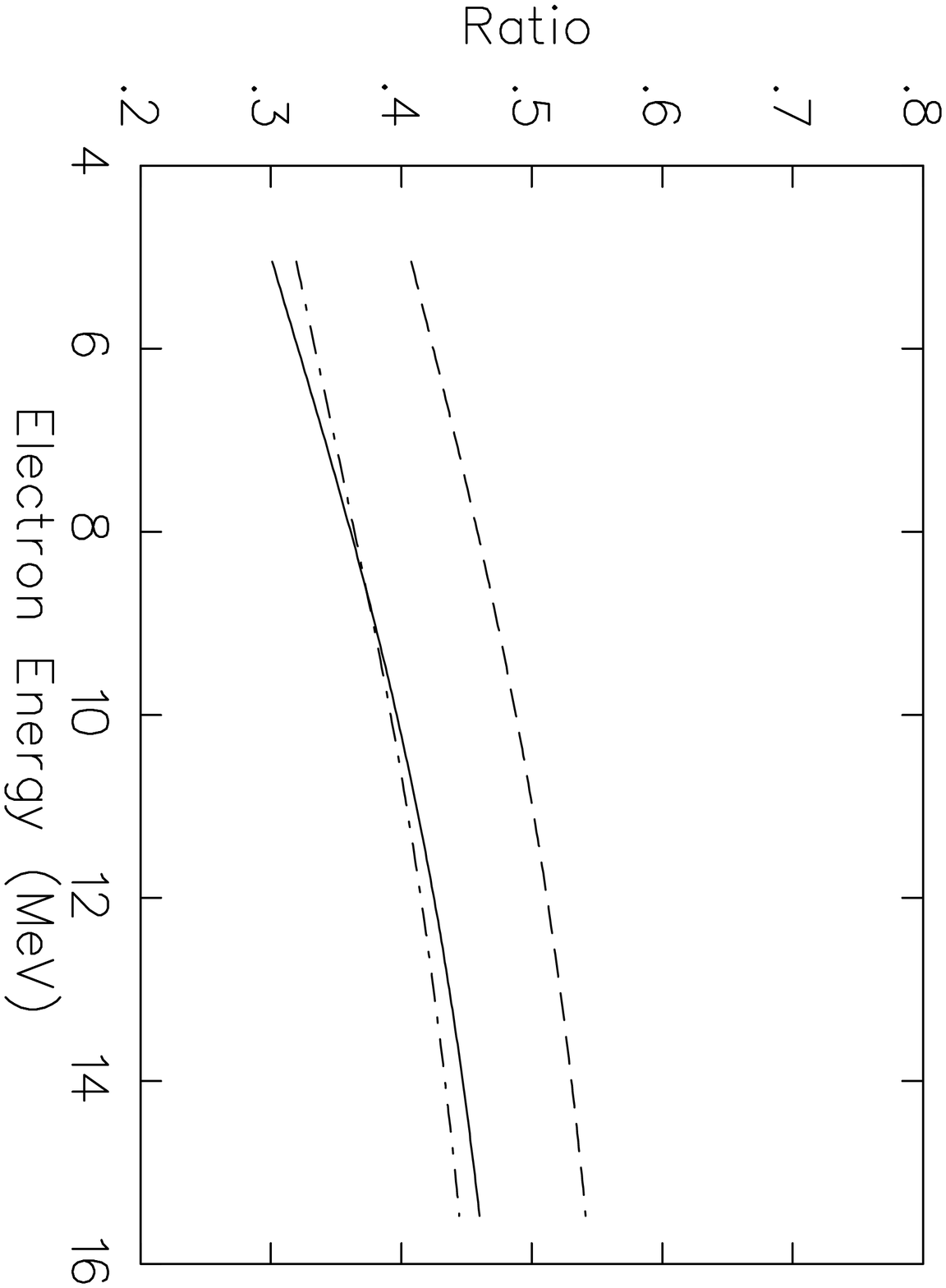}}
%\caption{}
\end{figure}

\begin{figure}
\epsfxsize=14cm
\centerline{\epsffile{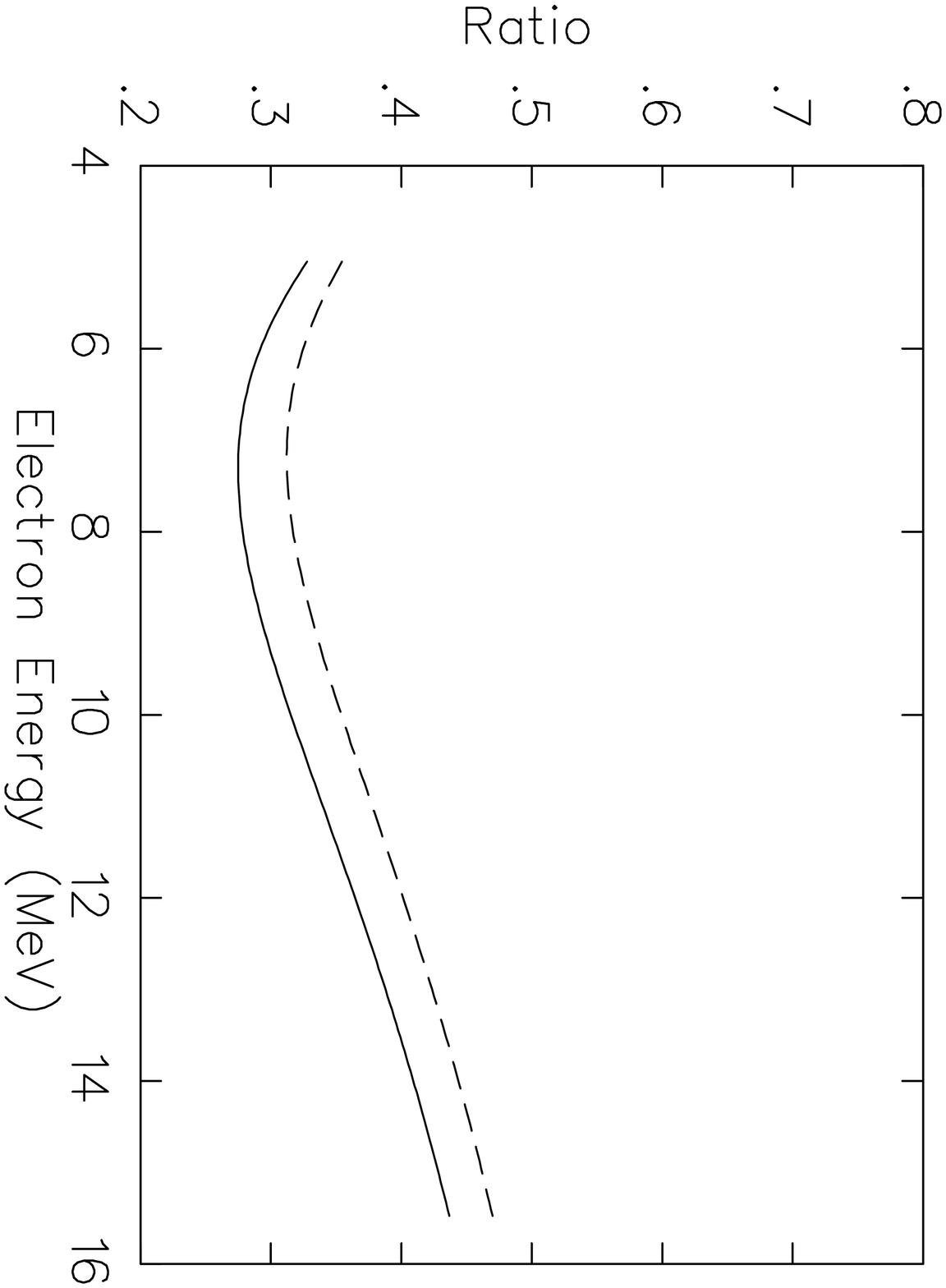}}
%\caption{}
\end{figure}

\end{document}